\documentclass[aip,graphicx]{revtex4-1}
\usepackage{amsmath}
\usepackage{amssymb}
\usepackage{mathrsfs}
\usepackage{natbib}
\usepackage{hyperref}
\usepackage{graphicx}
\usepackage{dcolumn}
\usepackage{bm}
\draft 
\usepackage{upgreek}
\usepackage{braket}
\renewcommand\bra[1]{{\langle{#1}|}}
\makeatletter
\renewcommand\ket[1]{%
  \@ifnextchar\bra{\k@t{#1}\!}{\k@t{#1}}%
}
\newcommand\k@t[1]{{|{#1}\rangle}}
\makeatother

\begin{document}


\title{Quantifying classical and quantum bounds for resolving closely spaced, non-interacting, simultaneously emitting dipole sources in optical microscopy} 

\author{Armine I. Dingilian}
\affiliation{Center for Biophysics and Quantitative Biology, University of Illinois Urbana-Champaign, Urbana, Illinois, USA}
\affiliation{NSF Science and Technology Center for Quantitative Cell Biology,University of Illinois Urbana-Champaign, Urbana, Illinois, USA}
\affiliation{Illinois Quantum Information Science and Technology Center, University of Illinois Urbana-Champaign, Urbana, Illinois, USA}

\author{Aarnah Kurella}
\affiliation{Department of Chemistry, University of Illinois Urbana-Champaign, Urbana, Illinois, USA}

\author{Cheyenne S. Mitchell}
\affiliation{Illinois Quantum Information Science and Technology Center, University of Illinois Urbana-Champaign, Urbana, Illinois, USA}
\affiliation{Department of Chemistry, University of Illinois Urbana-Champaign, Urbana, Illinois, USA}

\author{Dhananjay Dhruva}
\affiliation{Center for Biophysics and Quantitative Biology, University of Illinois Urbana-Champaign, Urbana, Illinois, USA}
\affiliation{NSF Science and Technology Center for Quantitative Cell Biology,University of Illinois Urbana-Champaign, Urbana, Illinois, USA}
\affiliation{Illinois Quantum Information Science and Technology Center, University of Illinois Urbana-Champaign, Urbana, Illinois, USA}

\author{David J. Durden}
\affiliation{Illinois Quantum Information Science and Technology Center, University of Illinois Urbana-Champaign, Urbana, Illinois, USA}
\affiliation{Department of Chemistry, University of Illinois Urbana-Champaign, Urbana, Illinois, USA}

\author{Mikael P. Backlund}
\email{mikaelb@illinois.edu}
\affiliation{Center for Biophysics and Quantitative Biology, University of Illinois Urbana-Champaign, Urbana, Illinois, USA}
\affiliation{NSF Science and Technology Center for Quantitative Cell Biology,University of Illinois Urbana-Champaign, Urbana, Illinois, USA}
\affiliation{Illinois Quantum Information Science and Technology Center, University of Illinois Urbana-Champaign, Urbana, Illinois, USA}
\affiliation{Department of Chemistry, University of Illinois Urbana-Champaign, Urbana, Illinois, USA}
\date{\today}

\begin{abstract}
Recent theoretical and experimental work has shown that the quantum Fisher information associated with estimating the separation between two optical point sources remains finite at small separations, effectively opening new routes to super-resolution imaging of simultaneously emitting sources. Most studies to date, however, implicitly invoke the scalar approximation, which is not appropriate in the context of high-numerical-aperture microscopy. Utilizing parameter estimation theory, here we consider the estimation of separation between two closely spaced dipole emitters, a commonly employed model for single-molecule optical beacons. We consider two limiting cases: one in which the orientations of the emitters are fixed and equal, and another in which both dipoles freely sample all of orientation space over the course of the measurement.  We quantify precision limits using quantum and classical variants of the Fisher information and Cram\'{e}r-Rao bound. In all cases, the vectorial nature of the emission complicates the analyses, but with appropriate filtering of the collected light in the azimuthal-radial polarization basis, a previously proposed scheme to saturate the quantum Fisher information via image inversion interferometry can be salvaged. 
\end{abstract}


\maketitle 

\section{Introduction}
Pioneering work in the 20th century by Helstrom and others extended concepts from classical detection and estimation theory to the realm of quantum metrology \cite{radhakrishna_rao_information_1945,mandel_fluctuations_1959,helstrom_minimum_1967,helstrom_estimation_1970,helstrom_resolution_1973,helstrom_quantum_1976,braunstein_statistical_1994,paris_quantum_2009}.
In 2016, Tsang and coworkers employed this view to reformulate the perennial problem of resolving mutually incoherent optical point sources spaced more closely than the diffraction limit. They found that the quantum information with respect to estimation of the separation between the sources was finite at all separations, indicating a possibility of extracting much more information from a suitable measurement than is laid bare by direct imaging \cite{tsang_quantum_2015,tsang_quantum_2016,tsang_quantum_2019,ang_quantum_2017}. In the years that followed, several experimental demonstrations of the basic concept have been published \cite{tang_fault-tolerant_2016,tham_beating_2017,paur_achieving_2016,yang_far-field_2016,donohue_quantum-limited_2018,paur_tempering_2018,larson_common-path_2019}. The theory has been extended to consider more general scenes, including source pairs with unequal brightness \cite{rehacek_multiparameter_2017,rehacek_optimal_2018,prasad_quantum_2020,bonsma-fisher_realistic_2019} and unknown centroid \cite{chrostowski_super-resolution_2017, grace_approaching_2020, parniak_beating_2018}, multiple point sources \cite{bisketzi_quantum_2019,lupo_quantum_2020} and extended objects \cite{tsang_subdiffraction_2017,tsang_subdiffraction_2018,zhou_modern_2019}. Variations on the emitted photons statistics have been considered including weak thermal light \cite{yang_fisher_2017,tsang_quantum_2011,hashemi_rafsanjani_quantum-enhanced_2017,lantz_quantum-enhanced_2017}, strong thermal light \cite{tsang_resolving_2019, yang_fisher_2017, nair_far-field_2016,lupo_ultimate_2016} and partial coherence \cite{larson_resurgence_2018,larson_common-path_2019,tsang_resurgence_2019,lee_surpassing_2019,wadood_superresolution_2019,hradil_quantum_2019}. Relevant to our purposes, the initial assumption of a Gaussian point spread function (PSF) has been relaxed to show the same idea works for any appropriately symmetric PSF \cite{kerviche_fundamental_2017,rehacek_optimal_2017}.
With the exception of our recent work \cite{mitchell_quantum-inspired_nodate}, all of the studies cited above implicitly invoke the scalar approximation, i.e. that the sources of radiation are effectively monopoles. For a net neutral collection of charges there is no monopole contribution to the radiation-- the lowest order contribution comes from the fluctuating electric dipole moment \cite{jackson_classical_1999}, which gives rise to a signature anisotropic distribution of radiation \cite{jackson_classical_1999,moerner_optical_1989,ha_single_1996,lukosz_light_1977,hellen_fluorescence_1987,born_principles_1999,polerecky_theory_2000,sepiol_single_1997,dickson_simultaneous_1998,bohmer_orientation_2003,lieb_single-molecule_2004,axelrod_fluorescence_2012,backer_extending_2014}. This distinction is inconsequential when the source is far enough away from the objective lens (as in telemetry and photography), but must be confronted when considering sources which are sufficiently close to the objective, as in high-numerical-aperture (NA) light microscopy \cite{ha_single_1996}. 
The importance of the dipolar nature of optical emission has long been appreciated in the field of single-molecule and super-resolution fluorescence microscopy, where high-NA collection is the norm \cite{hell_aberrations_1993,wiersma_comparison_1997,backlund_role_2014,karedla_simultaneous_2015}. The anisotropy of the fluorescence can be mined for orientational information \cite{ram_beyond_2006,backlund_simultaneous_2012,backer_bisected_2014,chao_fisher_2016}, and can lead to localization biases if ignored \cite{engelhardt_molecular_2011,lew_rotational_2013,lew_azimuthal_2014,backlund_removing_2016}. Recent work by Lew and coworkers have considered the full vectorial nature of emission in deriving the quantum Fisher information (QFI) associated with estimation of the position, orientation, and wobble of a single dipolar emitter \cite{zhang_fundamental_2019,ding_single-molecule_2020,ding_single-molecule_2021,zhang_single-molecule_2021,zhang_resolving_2022}, as well as the angular resolution of two dipole emitters located at the same position \cite{chen_resolving_2025}.
In the current work, we reconsider the quantum and classical bounds associated with the spatial resolution of simultaneously emitting optical sources, but now including the fully dipolar nature of the emission that is known to be important in high-NA microscopy. The basic findings remain largely the same, but optical designs to realize optimal measurements must be amended to account for the vectorial nature of the fields. In particular, we focus on the performance of an image inversion interferometer (III), which provides an elegant route to passive super-resolution in the case of monopole point sources by sorting the field by parity \cite{nair_interferometric_2016,wicker_characterisation_2009,weigel_investigation_2011,young_interferometric_2019,paur_reading_2019}. We find that for special, limiting cases of dipole orientation either fully parallel or perpendicular to the optical axis, the image inversion interferometer as previously proposed saturates the quantum Cram\'{e}r-Rao bound (QCRB); however, for arbitrary dipole orientations the method must be modified in order to approach these bounds. Namely, filtering the collected light in the azimuthal-radial polarization basis restores the super-resolving capability of the image inversion interferometer for arbitrary dipole orientations, providing a solution that can be readily implemented experimentally.
While passive super-resolution based on polarization-filtered image inversion interferometry is not expected to compete with established super-resolution fluorescence microscopy techniques \cite{balzarotti_nanometer_2017,gwosch_minflux_2020,huang_three-dimensional_2008} when imaging arbitrary, complicated scenes, it may provide an attractive alternative in cases where the simplicity of the scene is known a priori (e.g. that it consists of two loci) and the speedup allotted by foregoing sequential photoswitching is beneficial. In this work we ignore near-field interactions between dipoles, such as those that give rise to Förster resonance energy transfer (FRET), in order to focus the discussion on the effect of interest. It is well known that FRET is an invaluable reporter of intermolecular distance on length scales of the order of a few nanometers \cite{betzig_nobel_2015,moerner_nobel_2015,hell_nobel_2015,nelson_role_2018}. We leave it to future work to integrate the full array of near- and far-field effects that enable resolution below the diffraction limit. 

\section{Theory} \label{section_theory}
\begin{figure}[t]
    \centering
    \includegraphics{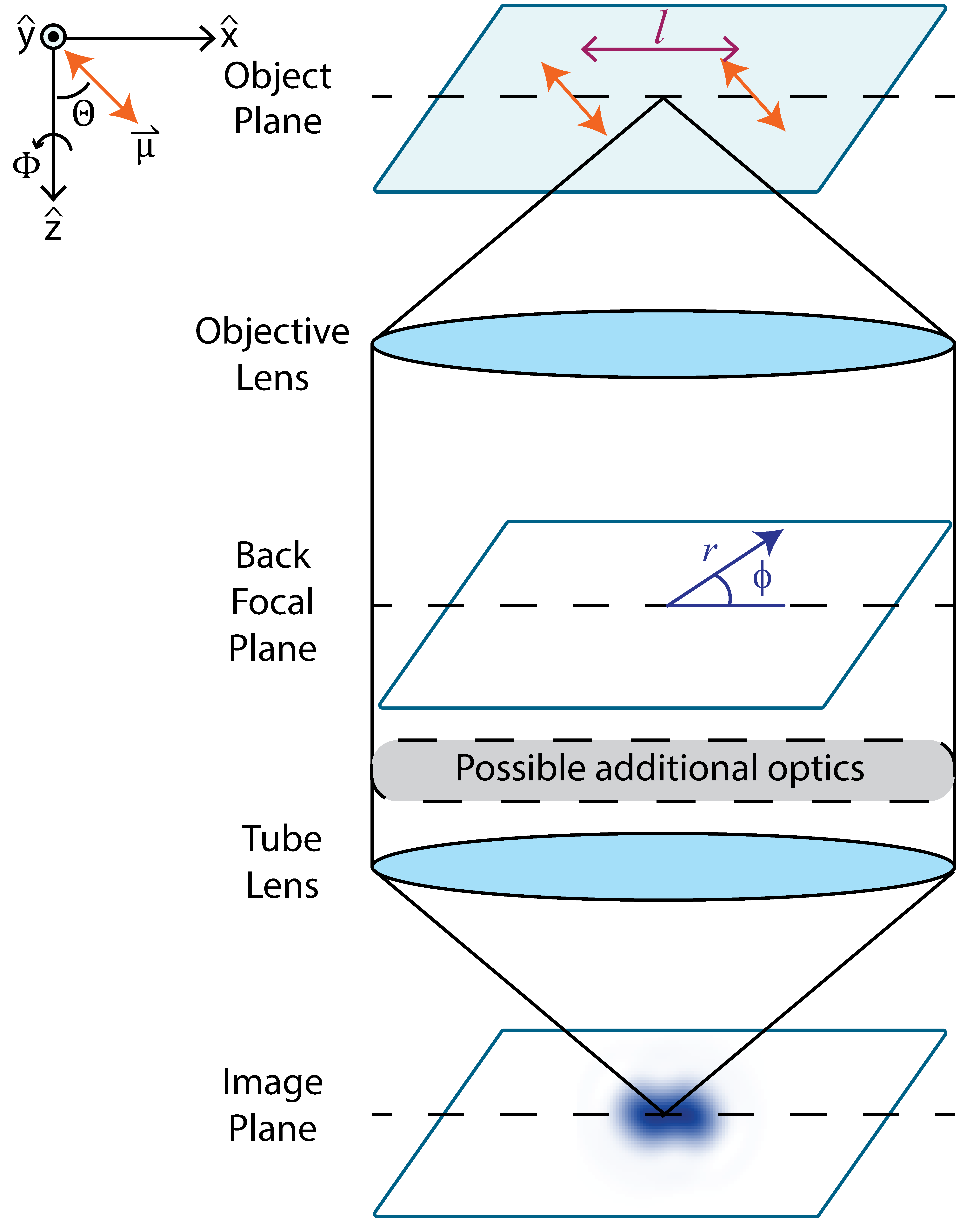}
    \caption{Overview schematic. We consider two non-interacting, mutually incoherent dipolar sources (orange arrows) of orientation $(\Theta,\Phi)$ symmetrically displaced from the optical axis and separated by a distance, $l$, to be estimated. The polar coordinates $(r,\phi)$ are defined at the back focal plane. Some of the measurements we consider require the placement of additional optical elements (e.g. beam splitters, polarizers, etc.) in the space between the back focal plane and the tube lens which forms the image(s) on the detector(s). Direct imaging is obtained without additional optics in this region. The optical axis is taken to be parallel to $\hat{z}$.}
    \label{fig:overview}
\end{figure}

We start by modeling two closely-spaced radiating dipole emitters embedded in a medium with index of refraction $n_1$, matched to that of the objective immersion medium (Fig. \ref{fig:overview}). This could, for example, correspond to two molecules embedded in a polymer supported on a coverslip, with index close to that of glass. In realistic experimental scenarios the orientation of the two dipoles will be unequal and unknown. In fact, this broken symmetry can aid the spatial resolution task \cite{lee_polarization-controlled_2012,hafi_fluorescence_2014}. The primary purpose of the current work, however, is to analyze how dipolar emission affects recently proposed quantum-inspired super-resolution concepts, so we choose to focus here on the limiting case in which the orientation of the two emitters is equal and known. We denote the polar and azimuthal orientations of both dipoles as $\Theta$ and $\Phi$, respectively (Fig. \ref{fig:overview} inset). We take the two dipoles to be symmetrically displaced from the origin of the object plane such that they are separated by a distance $l$ along the $\hat{x}$ direction. The separation distance $l$ is the unknown parameter to be estimated throughout this work. Over a sufficiently short interval the state of the electromagnetic field expressed at the back focal plane of the objective is a mixture of vacuum, $\rho_0$, and one-photon, $\rho_1$, states given by:
\begin{equation}
    \rho = (1-\epsilon)\rho_0 + \epsilon \rho_1,
\end{equation}
where $\epsilon$ is the probability of collecting a photon from the source pair. We take the total state of the field after $\mathscr{M}$ such intervals to be a direct product of these states, effectively ignoring correlations between subsequent photons. Throughout this manuscript we will report information metrics on a per-photon-collected basis. Anticipating this renormalization and noting that neither $\epsilon$ nor $\rho_0$ depend on $l$, we may as well reassign the state of the field to the one-photon state, $\rho \to \rho_1$. If we assume equal probabilities of the photon coming from either source, then:
\begin{equation}\label{eq_rhodef}
    \rho(\Theta,\Phi,l) = \frac{1}{2}\ket{\psi_+(\Theta,\Phi)}\bra{\psi_+(\Theta,\Phi)} + \frac{1}{2}\ket{\psi_-(\Theta,\Phi)}\bra{\psi_-(\Theta,\Phi)},
\end{equation}
where $\ket{\psi_\pm (\Theta,\Phi)}$ are the one-photon states resulting from emission of the source at positions $(x_o=\pm l/2,y_o=0,z_o=0)$ in the object plane. 

To construct $\ket{\psi_\pm (\Theta,\Phi)}$ we first consider the classical electric field due to a dipolar source of orientation $(\Theta,\Phi)$ located at position $x_o = \pm l/2$, as defined on the back focal plane of the objective \cite{backer_extending_2014}: 
\begin{equation} \label{eq_classicalfielddef}
    \mathbf{E}(r,\phi;\Theta,\Phi,\pm l/2) = e^{\mp i k_1 (r\cos\phi) l/2} \mathbf{G}(r,\phi)\cdot\hat{\mu}(\Theta,\Phi),
\end{equation}
where $(r,\phi)$ are the polar coordinates at the back focal plane (Fig. \ref{fig:overview}), defined such that $r$ is unitless and $\mathbf{E}(r)=0$ for $r>\text{NA}/n_1$, and $k_1=2\pi n_1/\lambda$ is the wavenumber in the immersion medium. Occasionally we will find it more convenient to work in terms of Cartesian coordinates at the back focal plane, $(x=r\cos\phi,y=r\sin\phi)$. The unit vector in the direction of the dipole is given by:
\begin{equation}
    \hat{\mu}(\Theta,\Phi) = \begin{pmatrix}
        \sin\Theta \cos\Phi \\ \sin\Theta \sin\Phi \\ \cos\Theta
    \end{pmatrix}
\end{equation}
and the Green's tensor is defined:
\begin{equation}
    \mathbf{G}(r,\phi) = \frac{E_0}{\left(1-r^2\right)^{1/4}} \begin{pmatrix}
        \left[\sin ^2\phi+\cos ^2\phi \sqrt{1-r^2}\right] & \left[\sin (2 \phi)\left(\sqrt{1-r^2}-1\right)/ 2\right]  & \left[-r \cos\phi\right] \\
\left[\sin (2 \phi)\left(\sqrt{1-r^2}-1\right) / 2\right] & \left[\cos ^2\phi+\sin ^2\phi \sqrt{1-r^2}\right] & \left[-r \sin \phi\right] \\
0 & 0 & 0
    \end{pmatrix},
\end{equation}
where $E_0$ is a complex constant we leave unspecified in anticipation of later normalization. The appropriate one-photon quantum states can then be defined:
\begin{equation} \label{eq_psipmxy}
    \ket{\psi_\pm(\Theta,\Phi)} = A(\Theta,\Phi) \iint \mathrm{d}x\mathrm{d}y \, \biggl\{\left[\hat{x}\cdot\mathbf{E}(r,\phi;\Theta,\Phi,\pm l /2)\right] \ket{x,y}_{\hat{x}} + \left[\hat{y}\cdot\mathbf{E}(r,\phi;\Theta,\Phi,\pm l /2)\right] \ket{x,y}_{\hat{y}}\biggr\},
\end{equation}
where $A(\Theta,\Phi)$ is a normalization constant and
\begin{subequations}
\begin{equation}
    \ket{x,y}_{\hat{x}} = \hat{a}^\dagger_{\hat{x}}(x,y)\ket{0},
\end{equation}
\begin{equation}
    \ket{x,y}_{\hat{y}} = \hat{a}^\dagger_{\hat{y}}(x,y)\ket{0},
\end{equation}
\end{subequations}
with $\ket{0}$ denoting the electromagnetic vacuum and $\hat{a}^\dagger_{\hat{x}}(x,y)$ [$\hat{a}^\dagger_{\hat{y}}(x,y)$] the creation operator for the mode localized to $(x,y)$ on the back focal plane with polarization $\hat{x}$ [$\hat{y}$]. Plugging Eq. (\ref{eq_psipmxy}) into Eq. (\ref{eq_rhodef}) gives an explicit expression for $\rho$.

The quantum Fisher information with respect to the parameter $l$ is \cite{helstrom_quantum_1976}
\begin{equation}
    \mathcal{K}(l;\Theta,\Phi) = \text{Tr}\left(\mathcal{L}_l^2\rho\right),
\end{equation}
where $\mathcal{L}_l$ is the symmetric logarithmic derivative (SLD) operator defined implicitly via:
\begin{equation}
    \partial_l\rho = \frac{\mathcal{L}_l\rho + \rho\mathcal{L}_l}{2}.
\end{equation}
An explicit expression for the SLD is given in the diagonal basis of $\rho$ by:
\begin{equation} \label{eq_SLDexplicit}
    \mathcal{L}_l = \sum_{\{k,k'|D_k+D_{k'}\neq0\}} \frac{2}{D_k + D_{k'}}\braket{k|\partial_l\rho|k'}\ket{k}\bra{k'},
\end{equation}
where
\begin{equation} \label{eq_rhodiagonal}
    \rho = \sum_k D_k \ket{k}\bra{k}.
\end{equation}
Thus evaluation of the QFI proceeds through diaganolization of $\rho$. The QCRB can then be readily computed:
\begin{equation}
    \sigma_\text{QCRB}^2(l;\Theta,\Phi) = \frac{1}{\mathcal{K}(l;\Theta,\Phi)}.
\end{equation}
The QFI provides an upper bound to the classical Fisher information (FI) of any microscope arrangement/choice of measurement. The QCRB, in turn, gives a lower bound to the classical Cram\'{e}r-Rao bound (CRB) of any such measurement. The gray boxes in each of Figs. \ref{fig:CRBxandzoriented}, \ref{fig:CRBoffangles}, and \ref{fig:CRBiso} are upper-bounded by $\sigma_\text{QCRB}$ for each indicated choice of $(\Theta,\Phi)$, meaning that a curve corresponding to $\sigma_\text{CRB}$ for any measurement scheme can at best touch the top of the gray box.

For comparison with the quantum bounds we consider several possible microscope arrangements, each culminating with spatially-resolved photon counting. Generically, let $I(x',y')$ denote the normalized intensity (i.e. the one-photon probability density) at position $(x',y')$ in the image plane. The one-parameter FI (per photon) is given by \cite{chao_fisher_2016}:
\begin{equation} \label{eq_CFI}
    \mathcal{J}(l) = \iint \mathrm{d}x' \mathrm{d}y' \, \frac{\left[\partial_l I(x',y') \right]^2}{I(x',y')},
\end{equation}
while the associated CRB is:
\begin{equation}
    \sigma^2_\text{CRB} = \frac{1}{\mathcal{J}},
\end{equation}
which provides a lower bound to the variance of any unbiased estimator of $l$. The most basic measurement scheme considered is ``direct imaging'', in which a tube lens placed one focal length downstream of the Fourier plane relays the collected light to a camera placed one focal length downstream of the lens (Fig. \ref{fig:overview}). The classical field on the image plane due to a dipole of orientation $(\Theta,\Phi)$ located in the object plane at $(x_o=\pm l/2,y_o=0,z_o=0)$ in this case is obtained by a scaled Fourier transform of the Fourier-plane field:
\begin{equation}
    \mathbf{E_\text{direct}}\left(x',y';\Theta,\Phi,\pm l/2\right) = C \iint \mathrm{d}x\mathrm{d}y \, \mathbf{E}\left(x,y;\Theta,\Phi,\pm l/2\right) \, e^{\frac{ik_1}{M}\left(xx' + yy'\right)},
\end{equation}
where $M$ is the magnification of the imaging system and $C$ is an unimportant constant that will be removed upon renormalization. The resulting (normalized) image is the incoherent sum of those due to both dipoles:
\begin{equation} \label{eq_directimg}
    I_\text{direct}(x',y';\Theta,\Phi,l) = \frac{\mathcal{N}(\Theta,\Phi)}{2}\left(\left| \mathbf{E_\text{direct}}\left(x',y';\Theta,\Phi,-l/2\right) \right|^2 + \left| \mathbf{E_\text{direct}}\left(x',y';\Theta,\Phi,l/2\right) \right|^2 \right),
\end{equation}
where $\mathcal{N}(\Theta,\Phi)$ is a normalization factor defined such that
\begin{equation}
    1=\iint\mathrm{d}x'\mathrm{d}y' \, I_\text{direct}\left(x',y';\Theta,\Phi,l\right).
\end{equation}
Calculation of $\mathcal{J}_\text{direct}(l;\Theta,\Phi)$ follows from plugging Eq. (\ref{eq_directimg}) into Eq. (\ref{eq_CFI}).

\begin{figure}[t]
    \centering
    \includegraphics{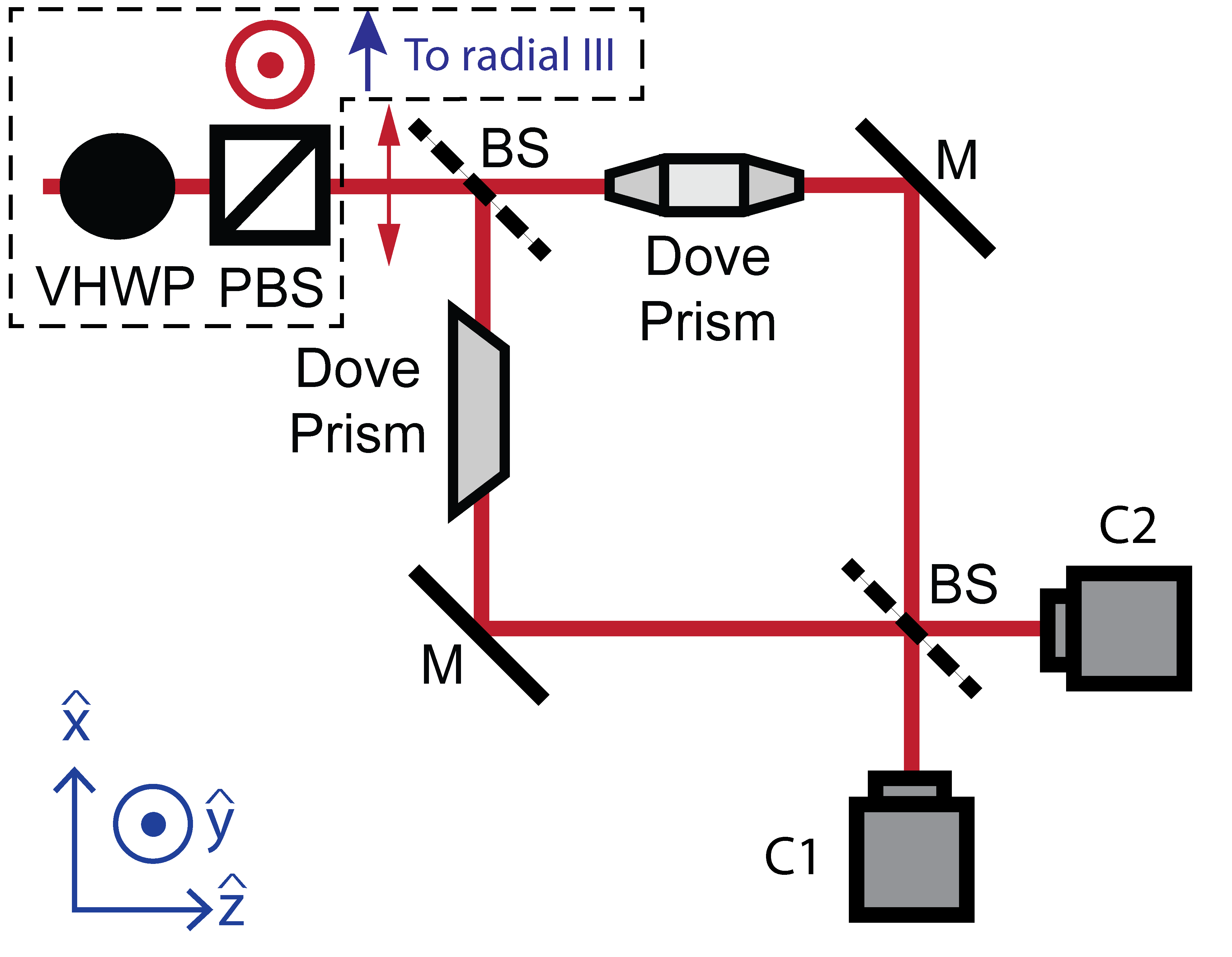}
    \caption{Schematic of the image inversion interferometer (III). Light collected by the microscope objective is injected from the left in this picture. VHWP: vortex half-wave plate, PBS: polarizing beam splitter, BS: 50:50 beam splitter, M: mirror, C1: camera 1, C2: camera 2. For clarity we have suppressed the tube lenses placed at both outputs of the second beam splitter which form the images at the detector planes. The portion enclosed in the dashed black box (VHWP and PBS) are removed for the unpolarized III measurement. The combination of the VWHP and PBS effectively resolves the collected field into its radially and azimuthally polarized components. In this work we consider measurements in which either or both of these polarized components are injected into a separate III. The III pictured here corresponds to the azimuthally polarized fraction, while the III handling the radially polarized part lies beyond the borders of the picture. The axes labeled in the lower left refer to the Cartesian coordinates just before the first beam splitter, as transformed from the object plane.}
    \label{fig:IIIsetup}
\end{figure}

Another measurement modality we consider is that of image inversion interferometry as initially proposed, without additional polarizing elements placed in the collection path (Fig. \ref{fig:IIIsetup}, excluding optical elements in dashed box). The first 50:50 beam splitter effects the transformation:
\begin{equation} \label{eq_BS1}
    \mathbf{E}(x,y;\Theta;\Phi,\pm l/2) \to \frac{i}{\sqrt{2}}\mathbf{E}(-x_R,y_R;\Theta,\Phi,\pm l/2) + \frac{1}{\sqrt{2}}\mathbf{E}(x_T,y_T;\Theta,\Phi,\pm l/2),
\end{equation}
where $(x_R,y_R)$ and $(x_T,y_T)$ denote the transverse coordinates in the reflected and transmitted modes, respectively. The Dove prism in the reflected arm leads to $x_R\to-x_R$, while the Dove prism in the transmitted arm gives $y_T\to-y_T$. After additional reflections and recombination on the second 50:50 beam splitter, the outputs of the interferometer are two fields given by:
\begin{subequations}
    \begin{equation}
        \mathbf{E^\text{(1)}}(x_1,y_1;\Theta,\Phi,\pm l/2)=\frac{1}{2}\biggl[ \mathbf{E}(-x_1,-y_1;\Theta,\Phi,\pm l/2) -  \mathbf{E}(x_1,y_1;\Theta,\Phi,\pm l/2) \biggr],
    \end{equation}
    \begin{equation}
        \mathbf{E^\text{(2)}}(x_2,y_2;\Theta,\Phi,\pm l/2)=\frac{i}{2}\biggl[ \mathbf{E}(-x_2,y_2;\Theta,\Phi,\pm l/2) +  \mathbf{E}(x_2,-y_2;\Theta,\Phi,\pm l/2) \biggr],
    \end{equation}
\end{subequations}
where $(x_1,y_1)$ and $(x_2,y_2)$ are the transvese coordinates at output ports 1 and 2 of the second beam splitter. A tube lens in both output ports results in the scaled Fourier transforms:
\begin{subequations}
    \begin{equation}
        \mathbf{E^\text{(1)}_\text{III}}(x'_1,y'_1;\Theta,\Phi,\pm l/2) = C \iint \mathrm{d}x_1\mathrm{d}y_1 \, \mathbf{E^\text{(1)}}\left(x_1,y_1;\Theta,\Phi,\pm l/2\right) \, e^{\frac{ik_1}{M}\left(x_1x'_1 + y_1y'_1\right)},
    \end{equation}
    \begin{equation}
        \mathbf{E^\text{(2)}_\text{III}}(x'_2,y'_2;\Theta,\Phi,\pm l/2) = C \iint \mathrm{d}x_2\mathrm{d}y_2 \, \mathbf{E^\text{(2)}}\left(x_2,y_2;\Theta,\Phi,\pm l/2\right) \, e^{\frac{ik_1}{M}\left(x_2x'_2 + y_2y'_2\right)}.
    \end{equation}
\end{subequations}
An image recorded in each output port is the incoherent sum of those due to both dipoles:
\begin{subequations} \label{eq_IIIunpolimg}
    \begin{equation}
        I^\text{(1)}_\text{III}(x'_1,y'_1;\Theta,\Phi,l) = \frac{\mathcal{N}(\Theta,\Phi)}{2}\left(\left| \mathbf{E^\text{(1)}_\text{III}}\left(x'_1,y'_1;\Theta,\Phi,-l/2\right) \right|^2 + \left| \mathbf{E^\text{(1)}_\text{III}}\left(x'_1,y'_1;\Theta,\Phi,l/2\right) \right|^2 \right),
    \end{equation}
        \begin{equation}
        I^\text{(2)}_\text{III}(x'_2,y'_2;\Theta,\Phi,l) = \frac{\mathcal{N}(\Theta,\Phi)}{2}\left(\left| \mathbf{E^\text{(2)}_\text{III}}\left(x'_2,y'_2;\Theta,\Phi,-l/2\right) \right|^2 + \left| \mathbf{E^\text{(2)}_\text{III}}\left(x'_2,y'_2;\Theta,\Phi,l/2\right) \right|^2 \right),
    \end{equation}
\end{subequations}
where the same factor $\mathcal{N}(\Theta,\Phi)$ as before ensures the normalization condition:
\begin{equation}
    1=\iint \mathrm{d}x'_1\mathrm{d}y'_1 \, I^\text{(1)}_\text{III}(x'_1,y'_1;\Theta,\Phi,l) + \iint \mathrm{d}x'_2\mathrm{d}y'_2 \, I^\text{(2)}_\text{III}(x'_2,y'_2;\Theta,\Phi,l).
\end{equation}
The FIs due to measurement in each output channel, $\mathcal{J}^\text{(1)}_\text{III}(l;\Theta,\Phi)$ and $\mathcal{J}^\text{(2)}_\text{III}(l;\Theta,\Phi)$, are obtained by plugging Eq. (\ref{eq_IIIunpolimg}) into Eq. (\ref{eq_CFI}). The total FI for the unpolarized III microscope is:
\begin{equation}
    \mathcal{J}_\text{III}(l;\Theta,\Phi) = \mathcal{J}^\text{(1)}_\text{III}(l;\Theta,\Phi) + \mathcal{J}^\text{(1)}_\text{III}(l;\Theta,\Phi). 
\end{equation}

Another microscope modality we consider is that of a specially polarized III in which the collected light is split into radially ($\hat{r}$)- and azimuthally ($\hat{\phi}$)-polarized components before entering separate image inversion interferometers (Fig. \ref{fig:IIIsetup}, including optical elements within dashed box). We also consider the case in which one or the other polarization is ignored after filtering, which was achieved experimentally recently in Ref. \cite{mitchell_quantum-inspired_nodate}. The polarization filtering is accomplished in two steps. First, the light passes through a vortex half-wave plate (VHWP) placed at a conjugate Fourier plane that effectively converts $\hat{r}$- into $\hat{x}$-polarized light and $\hat{\phi}$- into $(-\hat{y})$-polarized light \cite{backlund_removing_2016}. This transformation is described by a Jones matrix:
\begin{equation}
    \mathbf{J}(\phi) = \begin{pmatrix}
        \cos\phi & \sin\phi \\ \sin\phi & -\cos\phi
    \end{pmatrix}
\end{equation}
such that:
\begin{equation}
    \mathbf{E}(r,\phi;\Theta,\Phi,\pm l/2) \to \mathbf{J}(\phi) \cdot \mathbf{E}(r,\phi;\Theta,\Phi,\pm l/2).
\end{equation}
Next, a polarizing beam splitter (PBS) divides the light into two components we denote:
\begin{subequations}
    \begin{equation}
        E_{\hat{r}}(r_{\hat{r}},\phi_{\hat{r}};\Theta,\Phi,\pm l/2) = \hat{x}\cdot\mathbf{J}(\phi_{\hat{r}})\cdot\mathbf{E}(r_{\hat{r}},\phi_{\hat{r}};\Theta,\Phi,\pm l/2),
    \end{equation}
    \begin{equation}
        E_{\hat{\phi}}(r_{\hat{\phi}},\phi_{\hat{\phi}};\Theta,\Phi,\pm l/2) = \hat{y}\cdot\mathbf{J}(\phi_{\hat{\phi}})\cdot\mathbf{E}(r_{\hat{\phi}},\phi_{\hat{\phi}};\Theta,\Phi,\pm l/2),
    \end{equation}
\end{subequations}
where $(r_{\hat{r}},\phi_{\hat{r}})$ and $(r_{\hat{\phi}},\phi_{\hat{\phi}})$ are the polar transverse coordinates defined in the output channels of the PBS corresponding to light that was initially $\hat{r}$- and $\hat{\phi}$-polarized, respectively. Cartesian transverse coordinates are defined in both of these channels in the usual way such that $x_{\hat{r}}=r_{\hat{r}}\cos\phi_{\hat{r}}$, $y_{\hat{r}}=r_{\hat{r}}\sin\phi_{\hat{r}}$, $x_{\hat{\phi}}=r_{\hat{\phi}}\cos\phi_{\hat{\phi}}$, and $y_{\hat{\phi}}=r_{\hat{\phi}}\sin\phi_{\hat{\phi}}$. The differently-polarized fields output by the PBS then independently undergo a sequence of transformations analogous to Eqs. (\ref{eq_BS1}-\ref{eq_IIIunpolimg}), resulting in four output images we denote $I^\text{(1)}_{\hat{r}-\text{III}}(x'_{\hat{r},1},y'_{\hat{r},1};\Theta,\Phi,l)$, $I^\text{(2)}_{\hat{r}-\text{III}}(x'_{\hat{r},2},y'_{\hat{r},2};\Theta,\Phi,l)$, $I^\text{(1)}_{\hat{\phi}-\text{III}}(x'_{\hat{\phi},1},y'_{\hat{\phi},1};\Theta,\Phi,l)$, and $I^\text{(2)}_{\hat{\phi}-\text{III}}(x'_{\hat{\phi},2},y'_{\hat{\phi},2};\Theta,\Phi,l)$, obeying the overall normalization condition:
\begin{eqnarray}
    1 = && \iint \mathrm{d}x'_{\hat{r},1}\mathrm{d}y'_{\hat{r},1} \, I^\text{(1)}_{\hat{r}-\text{III}}(x'_{\hat{r},1},y'_{\hat{r},1};\Theta,\Phi,l) + \iint \mathrm{d}x'_{\hat{r},2}\mathrm{d}y'_{\hat{r},2} \, I^\text{(2)}_{\hat{r}-\text{III}}(x'_{\hat{r},2},y'_{\hat{r},2};\Theta,\Phi,l) \\ && + \iint \mathrm{d}x'_{\hat{\phi},1}\mathrm{d}y'_{\hat{\phi},1} \, I^\text{(1)}_{\hat{\phi}-\text{III}}(x'_{\hat{\phi},1},y'_{\hat{\phi},1};\Theta,\Phi,l) + \iint \mathrm{d}x'_{\hat{\phi},2}\mathrm{d}y'_{\hat{\phi},2} \, I^\text{(2)}_{\hat{\phi}-\text{III}}(x'_{\hat{\phi},2},y'_{\hat{\phi},2};\Theta,\Phi,l). \nonumber
\end{eqnarray}
Application of Eq. (\ref{eq_CFI}) gives the contribution to the FI from each of the four outputs, denoted $\mathcal{J}^\text{(1)}_{\hat{r}-\text{III}}(l;\Theta,\Phi)$, $\mathcal{J}^\text{(2)}_{\hat{r}-\text{III}}(l;\Theta,\Phi)$, $\mathcal{J}^\text{(1)}_{\hat{\phi}-\text{III}}(l;\Theta,\Phi)$, and $\mathcal{J}^\text{(2)}_{\hat{\phi}-\text{III}}(l;\Theta,\Phi)$. The total FI contributed by the $\hat{r}$-polarized light is
\begin{equation}
    \mathcal{J}_{\hat{r}-\text{III}}(l;\Theta,\Phi)=\mathcal{J}^\text{(1)}_{\hat{r}-\text{III}}(l;\Theta,\Phi) + \mathcal{J}^\text{(2)}_{\hat{r}-\text{III}}(l;\Theta,\Phi),
\end{equation}
while that contributed by the $\hat{\phi}$-polarized light is
\begin{equation}
    \mathcal{J}_{\hat{\phi}-\text{III}}(l;\Theta,\Phi)=\mathcal{J}^\text{(1)}_{\hat{\phi}-\text{III}}(l;\Theta,\Phi) + \mathcal{J}^\text{(2)}_{\hat{\phi}-\text{III}}(l;\Theta,\Phi).
\end{equation}
In the event that the $\hat{r}$-polarized ($\hat{\phi}$-polarized) signal is thrown away then the total recovered FI is given by $\mathcal{J}_{\hat{\phi}-\text{III}}$ ($\mathcal{J}_{\hat{r}-\text{III}}$). If both are retained then the total FI is:
\begin{equation}
    \mathcal{J}_{(\hat{r}+\hat{\phi})-\text{III}}(l;\Theta,\Phi)= \mathcal{J}_{\hat{r}-\text{III}}(l;\Theta,\Phi) + \mathcal{J}_{\hat{\phi}-\text{III}}(l;\Theta,\Phi).
\end{equation}

The above analysis assumes a dipole pair with fixed orientation $(\Theta,\Phi)$. The other limiting case we consider is that of a pair of isotropic emitters. Such an isotropic emitter might correspond to a single molecule tumbling freely and sufficiently fast such that it visits all of orientation space over the course of the measurement, or to an ensemble of uniformly randomly oriented dipoles located at the same position. It can easily be shown that the image of such an isotropic emitter can be obtained by an incoherent sum of images due to dipoles oriented along $\hat{x}$, $\hat{y}$, and $\hat{z}$. For a pair of isotropic emitters viewed with direct imaging, for instance, we have:
\begin{eqnarray}
    I_\text{direct}^{\text{(iso)}}(x',y';l) = \left(\frac{1}{2+\zeta}\right)\biggl[&&I_\text{direct}(x',y';\Theta=\pi/2,\Phi=0,l) \nonumber \\ &&+I_\text{direct}(x',y';\Theta=\pi/2,\Phi=\pi/2,l) \nonumber \\&&+\zeta I_\text{direct}(x',y';\Theta=0,\cdot,l) \biggr],
\end{eqnarray}
where $\zeta<1$ is the ratio of probabilities of collecting a photon from an emitting $\hat{z}$- oriented dipole to that of collecting a photon from an equally-emitting $\hat{x}$- oriented dipole. We leave $\Phi$ unspecified in the third term on the RHS of the above equations since it is not well-defined when $\Theta=0$. The ratio $\zeta$ can be computed from the elements of the Green's tensor via:
\begin{equation} \label{eq_zetadef}
    \zeta = \frac{\iint \mathrm{d}r\mathrm{d}\phi \, r^2 \left[ \left|\hat{x}\cdot\mathbf{G}(r,\phi)\cdot\hat{z}\right|^2 +  \left|\hat{y}\cdot\mathbf{G}(r,\phi)\cdot\hat{z}\right|^2\right] }{\iint \mathrm{d}r\mathrm{d}\phi \, r^2 \left[ \left|\hat{x}\cdot\mathbf{G}(r,\phi)\cdot\hat{x}\right|^2 +  \left|\hat{y}\cdot\mathbf{G}(r,\phi)\cdot\hat{x}\right|^2\right]}.
\end{equation}
For the images output by the unpolarized III we have:
\begin{eqnarray}
    I_\text{III}^{\text{(1,iso)}}(x'_1,y'_1;l) = \left(\frac{1}{2+\zeta}\right)\biggl[&&I_\text{III}^\text{(1)}(x'_1,y'_1;\Theta=\pi/2,\Phi=0,l) \nonumber \\ &&+I_\text{III}^\text{(1)}(x'_1,y'_1;\Theta=\pi/2,\Phi=\pi/2,l) \nonumber \\&&+\zeta I_\text{III}^\text{(1)}(x'_1,y'_1;\Theta=0,\cdot,l) \biggr]
\end{eqnarray}
and
\begin{eqnarray}
    I_\text{III}^{\text{(2,iso)}}(x'_2,y'_2;l) = \left(\frac{1}{2+\zeta}\right)\biggl[&&I_\text{III}^\text{(2)}(x'_2,y'_2;\Theta=\pi/2,\Phi=0,l) \nonumber \\ &&+I_\text{III}^\text{(2)}(x'_2,y'_2;\Theta=\pi/2,\Phi=\pi/2,l) \nonumber \\&&+\zeta I_\text{III}^\text{(2)}(x'_2,y'_2;\Theta=0,\cdot,l) \biggr].
\end{eqnarray}
Analogous expressions hold for all variants of the III. Likewise, for the sake of computing the QFI for a pair of isotropic emitters, the density operator is given by:
\begin{eqnarray}
    \rho_\text{iso}(l) = \left(\frac{1}{2+\zeta}\right)\biggl[&&\rho(\Theta=\pi/2,\Phi=0,l) \nonumber \\ &&+\rho(\Theta=\pi/2,\Phi=\pi/2,l) \nonumber \\&&+\zeta \rho(\Theta=0,\cdot,l) \biggr],
\end{eqnarray}
where the density operators on the RHS are defined as in Eq. (\ref{eq_rhodef}).
\section{Numerical methods}
Quantum and classical bounds were computed numerically in MATLAB. For all calculations we assumed a numerical aperture $\text{NA}=1.45$, magnification $M=100$, and an immersion medium matched to that of the sample with index $n_1=1.518$. We take the light to be quasimonochromatic with vacuum wavelength $\lambda = 670$ nm. To begin the calculation of the quantum bounds, we evaluated the classical fields at the back focal plane according to Eq. (\ref{eq_classicalfielddef}), sampled finely on a $2049\times2049$ grid such that the circular support corresponded to about 60\% of the total area. Normalization yields the states $\ket{\psi_{\pm}(\Theta,\Phi)}$ represented in a discretized approximation of the continuous basis $\left\{\{\ket{x,y}_{\hat{x}}\},\{\ket{x,y}_{\hat{y}}\}\right\}$. In principle, this can be used to approximately represent $\rho(\Theta,\Phi,l)$ as a $2049^2\times2049^2$ matrix, which can in turn be diagonalized to compute the SLD according to Eqs. (\ref{eq_rhodiagonal}) and (\ref{eq_SLDexplicit}). In practice, however, this matrix tends to be poorly conditioned and therefore its diagonalization prone to numerical errors. Instead, prior to diagonalization we first transformed $\ket{\psi_\pm(\Theta,\Phi)}$ into a suitable discrete basis. Given the circular support $r\leq\text{NA}/n_1$, the Zernike polynomials are a natural choice, defined by:
\begin{equation}
    Z_n^m\left(\frac{n_1 r}{\text{NA}},\phi\right) = \begin{cases}
        R_n^m\left(\frac{n_1 r}{\text{NA}}\right)\cos(m\phi), \quad m \geq 0 \\ R_n^m\left(\frac{n_1 r}{\text{NA}}\right)\sin(m\phi), \quad m<0
    \end{cases}
\end{equation}
for all nonnegative integers $n$ and $m\in\{-n,-n+2,...,n-2,n\}$. The radial polynomials are given by:
\begin{equation}
    R_n^m\left(\frac{n_1 r}{\text{NA}}\right) = 
        \sum_{k=0}^{\frac{n-m}{2}} \frac{(-1)^k(n-k)!}{k! \left(\frac{n+m}{2}-k\right)! \left(\frac{n-m}{2}-k\right)!} \left(\frac{n_1 r}{\text{NA}}\right)^{n-2k} \text{circ}\left(\frac{n_1 r}{\text{NA}}\right)
\end{equation}
with $\text{circ}(u)=1$ for $u\leq1$ and 0 otherwise. Expansion of $\ket{\psi_{\pm}(\Theta,\Phi}$ in this basis proceeds by numerical evaluation of the coefficients $_{\hat{x}}\braket{Z_n^m|\psi_{\pm}(\Theta,\Phi)}$ and $_{\hat{y}}\braket{Z_n^m|\psi_{\pm}(\Theta,\Phi)}$, where
\begin{subequations}
    \begin{equation}
        \ket{Z_n^m}_{\hat{x}} = \iint \mathrm{d}x\mathrm{d}y \, Z_n^m\bigl(r(x,y),\phi(x,y)\bigr)\ket{x,y}_{\hat{x}},
    \end{equation}
    \begin{equation}
        \ket{Z_n^m}_{\hat{y}} = \iint \mathrm{d}x\mathrm{d}y \, Z_n^m\bigl(r(x,y),\phi(x,y)\bigr)\ket{x,y}_{\hat{y}}.
    \end{equation}
\end{subequations}
We truncated the expansion into Zernike modes at $n = 8$, resulting in a density matrix of dimension $90\times90$, which was then diagonalized to compute the SLD and, in turn, the QFI and QCRB.

To calculate classical bounds, we again began by computing the fields $\mathbf{E}(r,\phi;\Theta,\Phi,\pm l/2)$ defined at the back focal plane. These fields were then transformed according to the equations detailed in the previous section. Finally, the image-plane fields were computed numerically via a two-dimensional fast Fourier transform. Sampling of the field on the back focal plane was chosen such that the final images were sampled on a grid with pixels of size 5 nm $\times$ 5 nm projected back to object space. With the images in hand, the FI and CRB were then calculated numerically.

\section{Results and Discussion}
\subsection{Fixed-orientation dipole sources}

\begin{figure}
    \centering
    \includegraphics{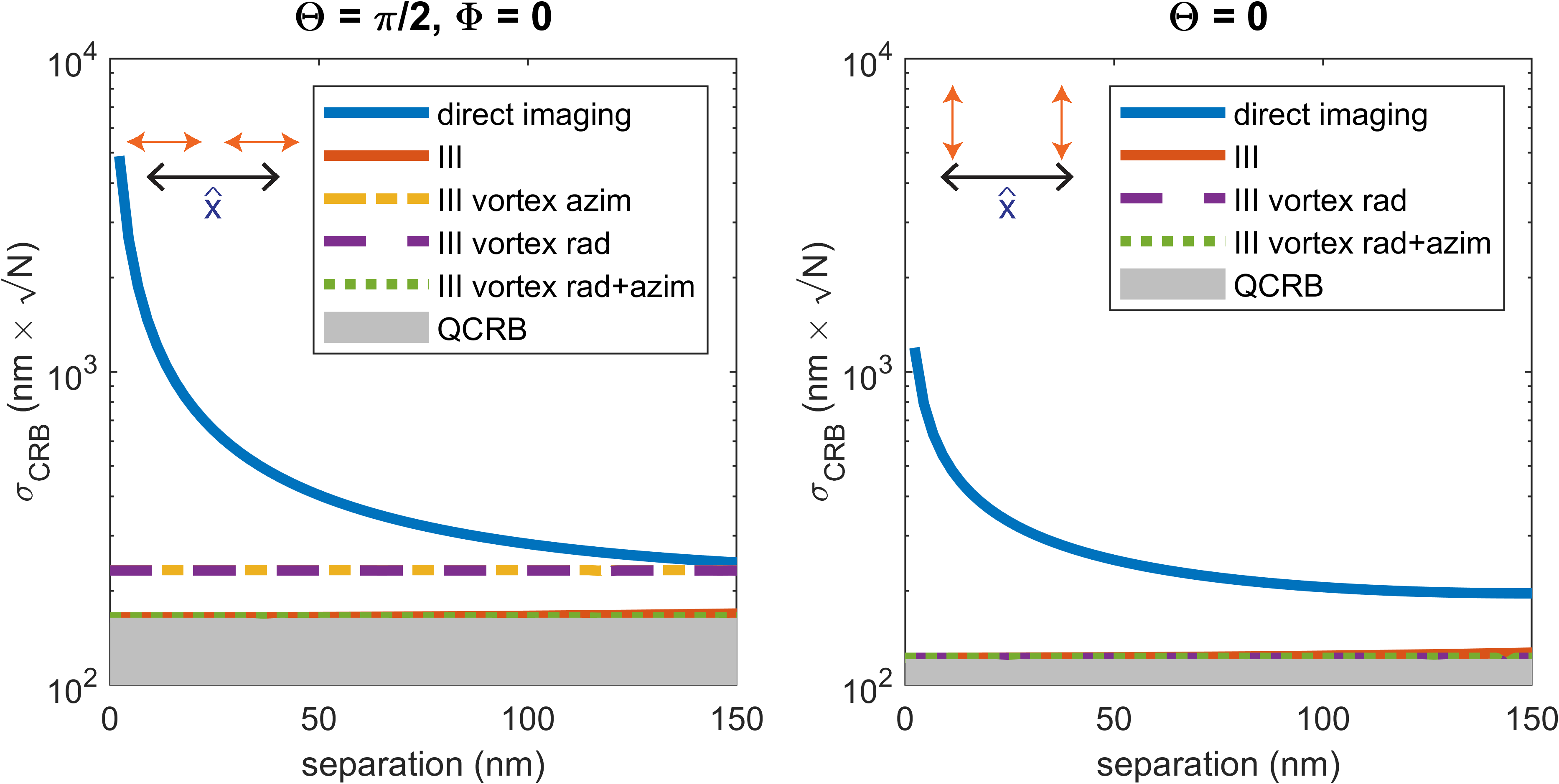}
    \caption{Computed CRBs vs. separation for the special cases $(\Theta=\pi/2,\Phi=0)$ (left) and $\Theta=0$ (right). The gray boxes are bounded above by the QCRBs. $N$ is the number of photons collected. Labels in the legends refer to direct imaging (blue solid), unpolarized III (red solid), $\hat{\phi}$-polarized III (yellow dashed), $\hat{r}$-polarized III (purple dashed), and the combination of separate $\hat{\phi}$- and $\hat{r}$-polarized III measurements (green dotted). The gold dashed line does not appear in the right plot because all of the light collected in this case is $\hat{r}$-polarized.}
    \label{fig:CRBxandzoriented}
\end{figure}

\begin{figure}
    \centering
    \includegraphics[width = \textwidth]{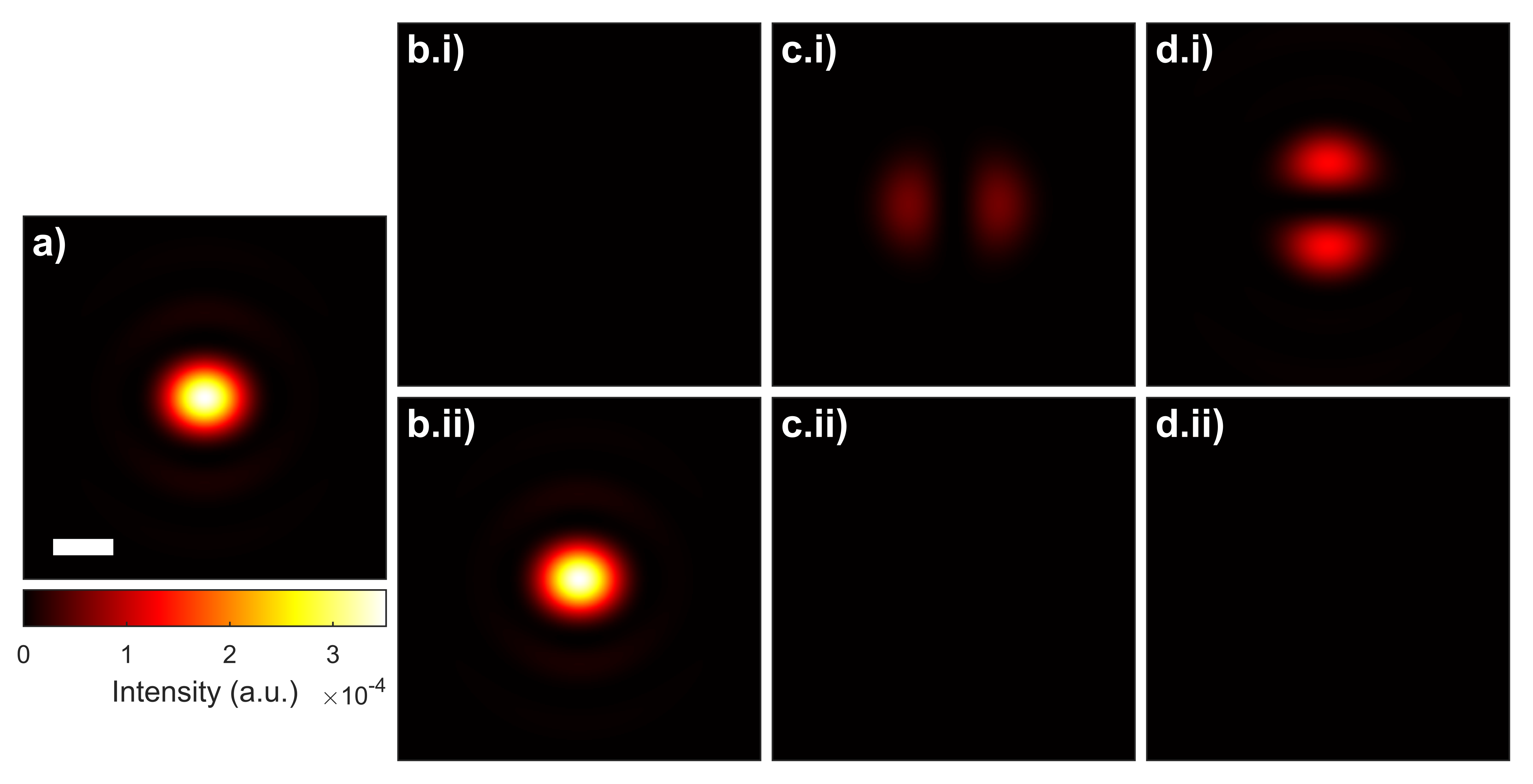}
    \caption{High-resolution simulated images for $(\Theta=\pi/2,\Phi=0)$ and fixed $l\approx10$ nm. (a) Direct imaging. (b) Unpolarized III Outputs 1 (b.i) and 2 (b.ii). (c) $\hat{r}$-polarized III Outputs 1 (c.i) and 2 (c.ii). (d) $\hat{\phi}$-polarized III Outputs 1 (d.i) and 2 (d.ii). Scale bar: 250 nm.}
    \label{fig:simimagesthetapi2phi0}
\end{figure}

Results for a pair of dipole emitters with orientation parallel to $\hat{x}$ (i.e. $\Theta = \pi/2$, $\Phi=0$) are depicted in the left panel of Fig. \ref{fig:CRBxandzoriented}. ``Rayleigh's Curse'' \cite{tsang_quantum_2016} (i.e. the divergence of the CRB as $l\to0$) is evident for direct imaging at small separations, where $\sigma_\text{CRB}^{\text{(direct)}} \gg \sigma_\text{QCRB}$. In this case, the unpolarized III measurement subverts Rayleigh's Curse and saturates the quantum bound. Thus, for the special case of $(\Theta=\pi/2,\Phi=0)$ the III setup as initially proposed \cite{nair_far-field_2016,tang_fault-tolerant_2016,nair_interferometric_2016} works as expected, without the need for additional polarization filtering. Some intuition can be gleaned by inspection of the simulated images in Fig. \ref{fig:simimagesthetapi2phi0}b corresponding to a separation of about 10 nm (actual separation here is 8.99 nm due to a quirk of sampling). The III routes nearly all of the collected light to Output 2. The small fraction of light shunted to Output 1 increases as separation increases, and the super-resolving effect emerges from the ability to sit on this dark fringe. The observation that nearly all of the light exits Output 2 (the symmetric port) at small separations is rationalized by examining the inversion symmetry of $\mathbf{E}(r,\phi;\pi/2,0,0)$, namely that
\begin{equation}
    \mathbf{E}(r,\pi-\phi;\pi/2,0,0) = \mathbf{E}(r,\phi;\pi/2,0,0).
\end{equation}
In fact, this symmetry holds for any $\Phi$ when $\Theta=\pi/2$, i.e. when the dipoles are perpendicular to the optical axis.

\begin{figure}
    \centering
    \includegraphics[width=\linewidth]{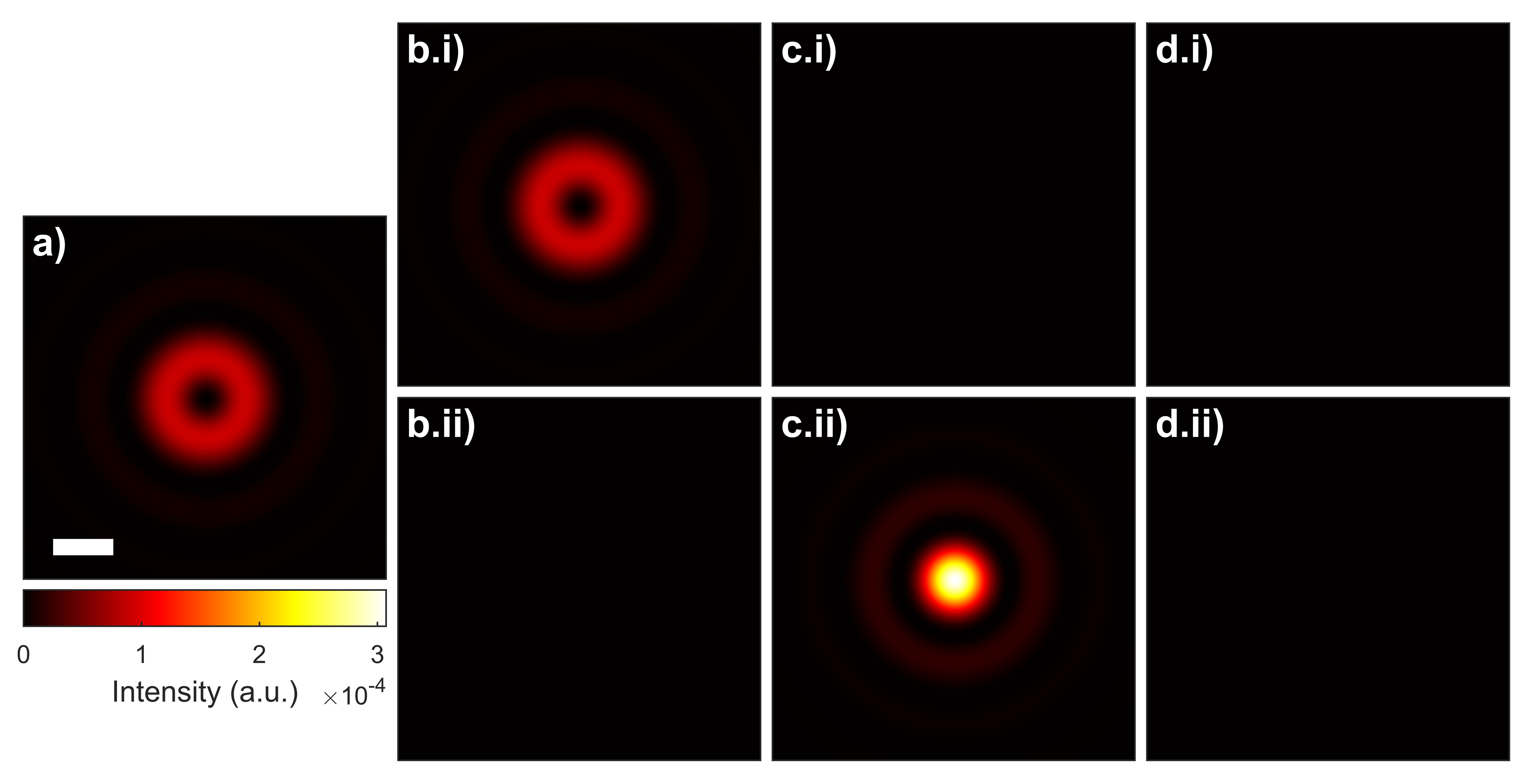}
    \caption{High-resolution simulated images for $\Theta=0$ and fixed $l\approx10$ nm. (a) Direct imaging. (b) Unpolarized III Outputs 1 (b.i) and 2 (b.ii). (c) $\hat{r}$-polarized III Outputs 1 (c.i) and 2 (c.ii). (d) $\hat{\phi}$-polarized III Outputs 1 (d.i) and 2 (d.ii). Scale bar: 250 nm.}
    \label{fig:simimagestheta0phi0}
\end{figure}

The right panel of Fig. \ref{fig:CRBxandzoriented} shows another special case for which the unpolarized III saturates the quantum bound. Here the dipoles are parallel to the optical ($\hat{z}$) axis, such that $\Theta=0$. Inspection of the simulated images in Fig. \ref{fig:simimagestheta0phi0}b shows that at small separations the unpolarized III shunts almost all of the light to Output 1 in this case, a result of the antisymmetry of the relevant field:
\begin{equation}
    \mathbf{E}(r,\pi-\phi;0,\cdot,0) = -\mathbf{E}(r,\phi;0,\cdot,0).
\end{equation}
As an aside, we note that the CRB for direct imaging for $\hat{z}$-oriented sources is slightly improved relative to that of $\hat{x}$-oriented sources. This likely can be ascribed to a recently reported phenomenon by which PSFs exhibiting isolated zeros lead to improved scaling of the resolution at small separations \cite{paur_reading_2019}. The comparison between the two orientations is complicated by the choice to normalize based on photons collected (rather than emitted). It is known that a $\hat{z}$-oriented dipole will generally appear dimmer than an equally-emitting $\hat{x}$- oriented dipole, leading to $\zeta<1$ in Eq. (\ref{eq_zetadef}). Assessing this penalty makes the modest resolution enhancement due to this effect more modest still.

\begin{figure}
    \centering
    \includegraphics{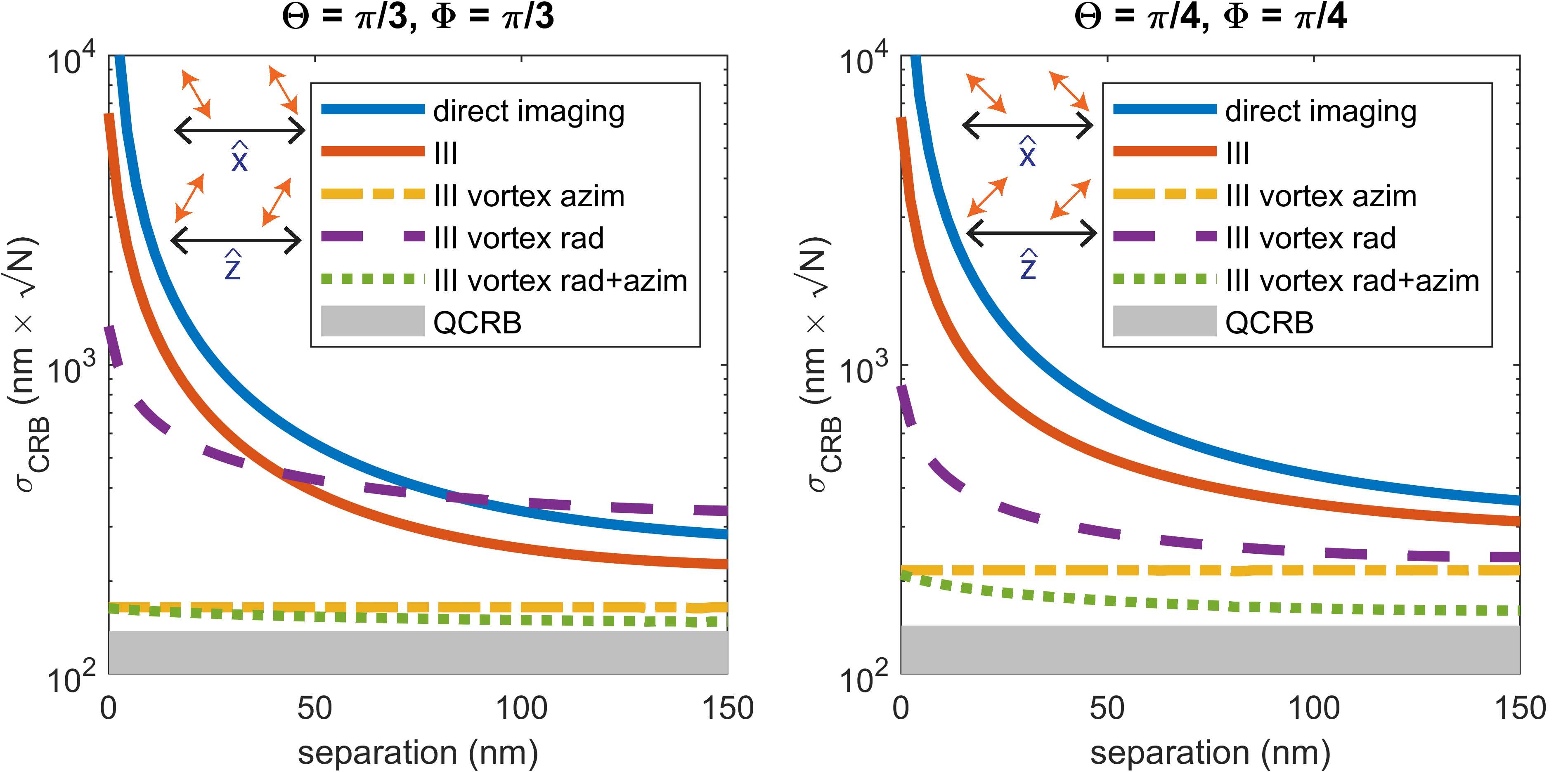}
    \caption{Computed CRBs vs. separation for the special cases $(\Theta=\pi/3,\Phi=\pi/3)$ (left) and $(\Theta=\pi/4,\Phi=\pi/4)$ (right). The gray boxes are bounded above by the QCRBs. $N$ is the number of photons collected. Labels in the legends refer to direct imaging (blue solid), unpolarized III (red solid), $\hat{\phi}$-polarized III (yellow dashed), $\hat{r}$-polarized III (purple dashed), and the combination of separate $\hat{\phi}$- and $\hat{r}$-polarized III measurements (green dotted).}
    \label{fig:CRBoffangles}
\end{figure}

\begin{figure}
    \centering
    \includegraphics[width = \textwidth]{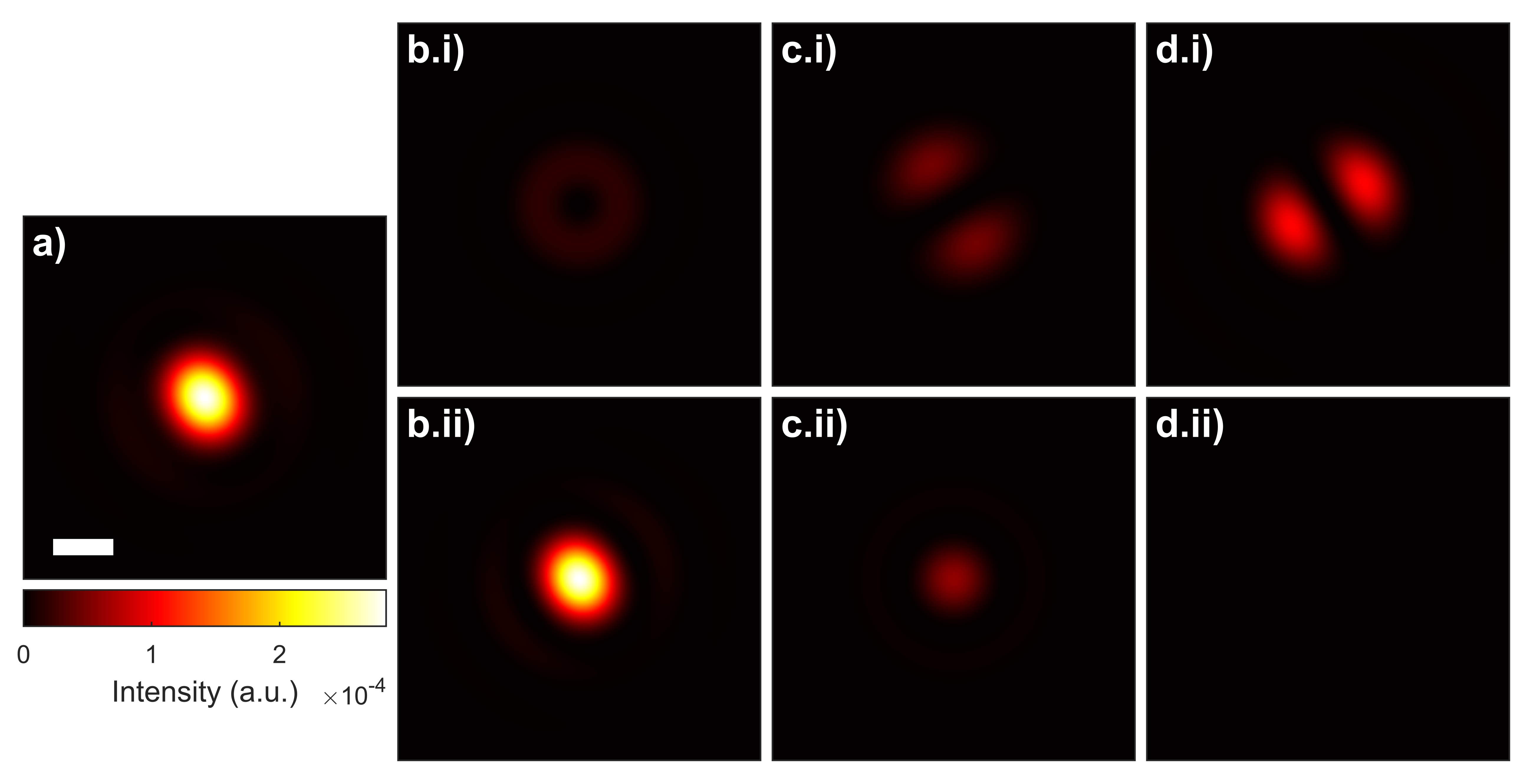}
    \caption{High-resolution simulated images for $(\Theta=\pi/3,\Phi=\pi/3)$ and fixed $l\approx10$ nm. (a) Direct imaging. (b) Unpolarized III Outputs 1 (b.i) and 2 (b.ii). (c) $\hat{r}$-polarized III Outputs 1 (c.i) and 2 (c.ii). (d) $\hat{\phi}$-polarized III Outputs 1 (d.i) and 2 (d.ii). Scale bar: 250 nm}
    \label{fig:simimagesthetapi3phipi3}
\end{figure}

In the extreme cases of dipoles oriented either parallel ($\Theta=0$) or perpendicular ($\Theta=\pi/2$) to the optical axis, the unpolarized III measurement requires no modification to achieve quantum-inspired super-resolution. This is not true, however, for intermediate values of $\Theta$. Figure \ref{fig:CRBoffangles} shows results for two such orientations with $\Theta = \pi/3$ and $\Theta=\pi/4$. In both cases the CRB of the unpolarized III measurement is only slightly better than that of direct imaging. Here the field $\mathbf{E}(r,\phi;\Theta,\Phi,0)$ contains both a symmetric and an antisymmetric component, and therefore is overall asymmetric with respect to inversion. As a result, the unpolarized III cannot produce efficient nulling in either output port as $l\to0$ (Fig. \ref{fig:simimagesthetapi3phipi3}b). From previous work \cite{lew_azimuthal_2014,backlund_removing_2016} it is known that for any $(\Theta,\Phi)$, resolving $\mathbf{E}(r,\phi;\Theta,\Phi,0)$ into the $\hat{r}-\hat{\phi}$ polarization basis yields a $\hat{\phi}$ component with definite parity. Such polarization filtering can be realized in practice with a vortex half-wave plate and a linear polarizer (Fig. \ref{fig:IIIsetup}), as described in Section \ref{section_theory}. Isolating the $\hat{\phi}$ component before injecting into the III salvages the ability to effectively null one of the output ports as $l\to0$ (Fig. \ref{fig:simimagesthetapi3phipi3}d). As can be seen from the gold dashed lines in Fig. \ref{fig:CRBoffangles}, this measurement recovers nearly all of the super-resolution promised by the QCRB, leaving only a small gap associated with the cost of throwing away the $\hat{r}$-polarized component of the collected light. Of course, this portion of the light can be processed and measured as well-- one possibility is to reroute to a second III (Fig. \ref{fig:IIIsetup}). The purple and green lines in Fig. \ref{fig:CRBoffangles} show that this approach helps to close the gap at moderate subdiffraction separations, but contributes no additional information at vanishing $l$. The main benefit of the dual-III measurement is that one does not sacrifice the ability to saturate the QCRB in the limiting cases of $\Theta=\pi/2$ and $\Theta=0$ (Fig. \ref{fig:CRBxandzoriented}). For $\Theta=\pi/2$ the two polarized III measurements are equally informative. For $\Theta=0$ the collected light is entirely $\hat{r}$-polarized and so the $\hat{\phi}$-polarized measurement contributes nothing to the task.

\begin{figure}
    \centering
    \includegraphics[width = 0.5\textwidth]{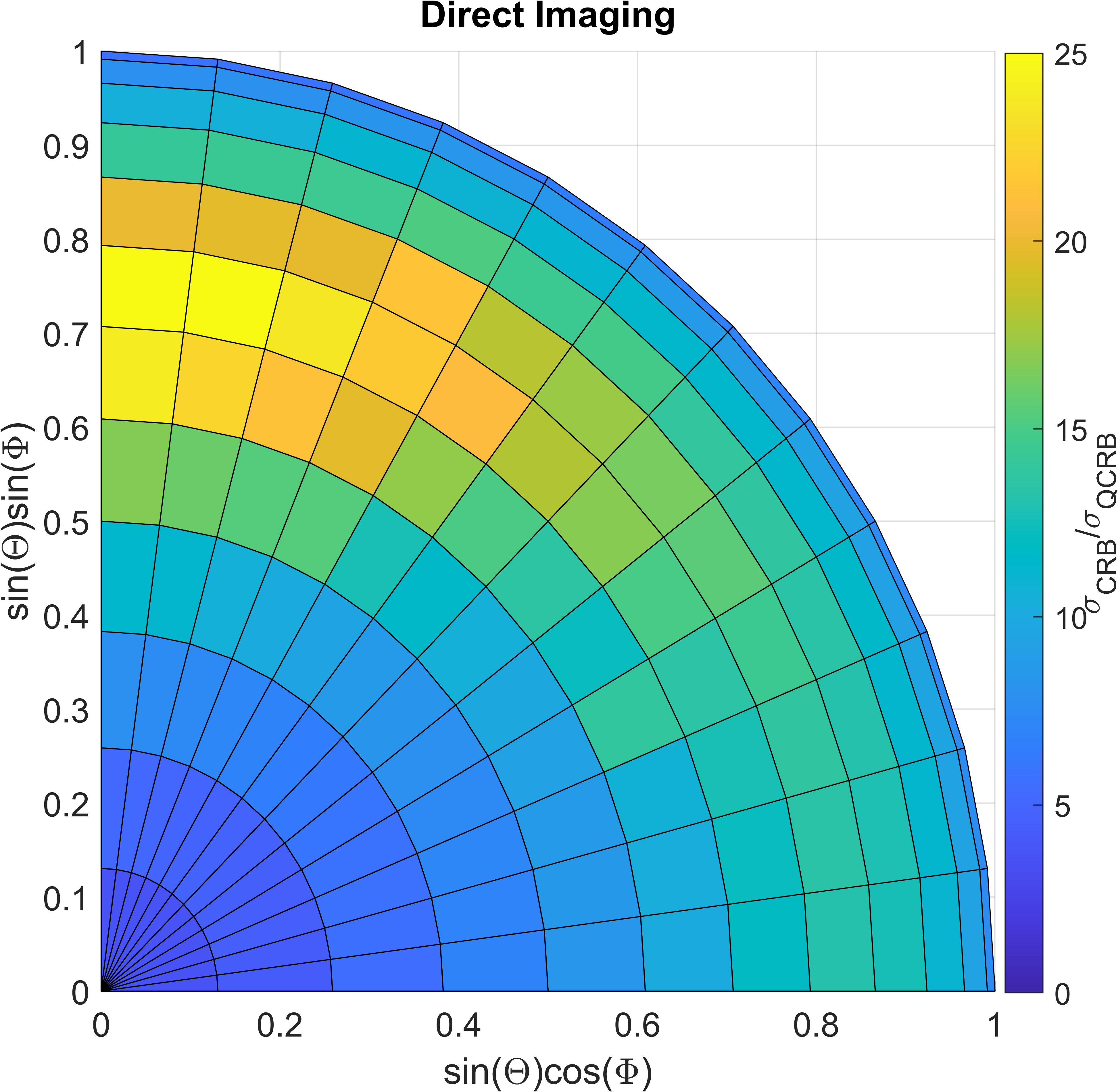}
    \caption{The ratio $\sigma_\text{CRB}/\sigma_\text{QCRB}$ for direct imaging at fixed $l\approx10$ nm and $\Theta \in [0,\pi/2]$, $\Phi \in [0,\pi/2]$ sampled at intervals of $\pi/12$.}
    \label{fig:polarplotdirect}
\end{figure}

\begin{figure}
    \centering
    \includegraphics[width = 0.5\textwidth]{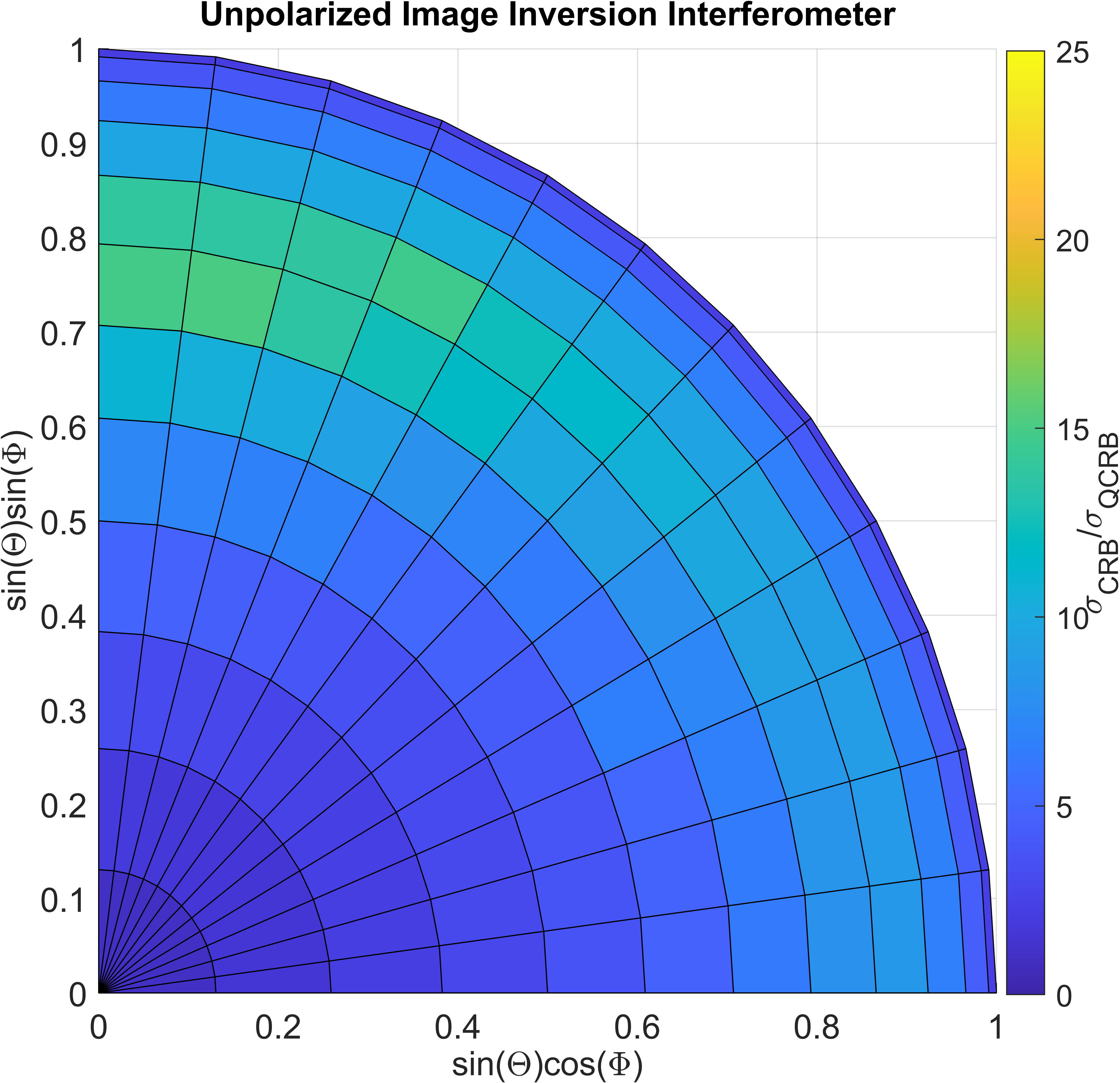}
    \caption{The ratio $\sigma_\text{CRB}/\sigma_\text{QCRB}$ for the unpolarized III measurement at fixed $l\approx10$ nm and $\Theta \in [0,\pi/2]$, $\Phi \in [0,\pi/2]$ sampled at intervals of $\pi/12$.}
    \label{fig:polarplotunpolIII}
\end{figure}

\begin{figure}
    \centering
    \includegraphics[width = 0.5\textwidth]{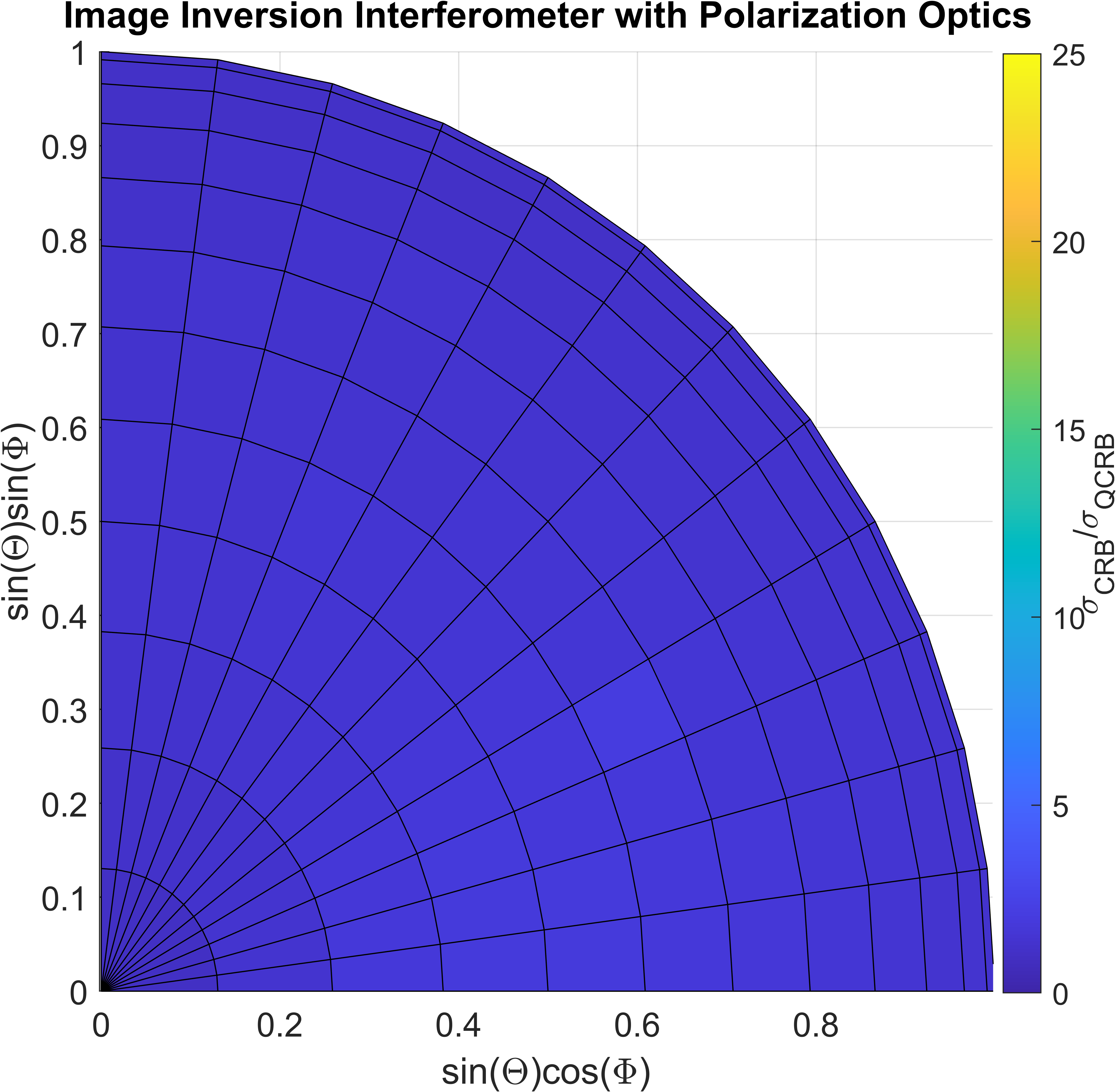}
    \caption{The ratio $\sigma_\text{CRB}/\sigma_\text{QCRB}$ for the measurement combining $\hat{r}$- and $\hat{\phi}$-polarized III at fixed $l\approx10$ nm and $\Theta \in [0,\pi/2]$, $\Phi \in [0,\pi/2]$ sampled at intervals of $\pi/12$.}
    \label{fig:polarplotpolIII}
\end{figure}

Comparisons of classical and quantum bounds across the entire range of $(\Theta,\Phi)$ are depicted in Figs. \ref{fig:polarplotdirect}-\ref{fig:polarplotpolIII} for fixed $l\approx10$ nm. The color map in these polar plots encodes the ratio $\sigma_\text{CRB}/\sigma_\text{QCRB}$ for the various measurements considered and is fixed in scale across all three figures. Figure \ref{fig:polarplotdirect} shows that this ratio hovers around an order of magnitude for most of orientation space. The fact that the ratio is slightly larger for $\Phi$ near $\pi/2$ than it is for $\Phi$ near 0 is due to our choice to make the direction of separation parallel to $\hat{x}$. Rotating the direction of separation yields a rotated version of the plot; averaging over all separation directions, as would be appropriate if no prior information about separation direction is available, symmetrizes the plot. Figure \ref{fig:polarplotunpolIII} shows that an unpolarized III measurement reduces $\sigma_\text{CRB}/\sigma_\text{QCRB}$ to the order of unity for some $(\Theta,\Phi)$, but leaves it around 10 for others. Finally, Fig. \ref{fig:polarplotpolIII} shows that the measurement in which the collected light is resolved into the $\hat{r}-\hat{\phi}$ basis and both components are directed into separate III setups reduces $\sigma_\text{CRB}/\sigma_\text{QCRB}$ to within a factor of $\sim$2 across all of orientation space.

\subsection{Isotropic sources}

\begin{figure}
    \centering
    \includegraphics{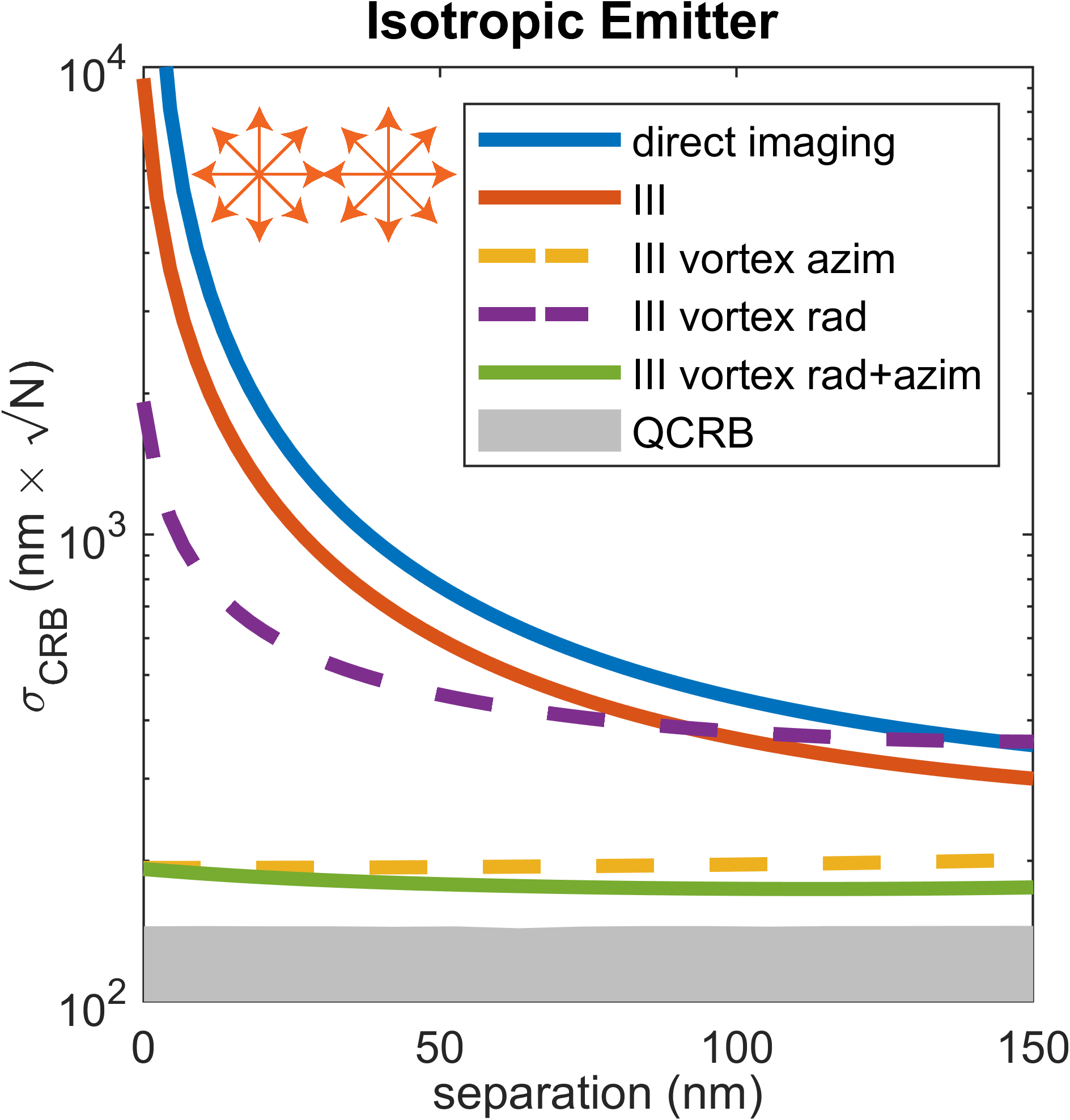}
    \caption{Computed CRBs vs. separation for a pair of isotropic emitters. The gray boxes are bounded above by the QCRBs. $N$ is the number of photons collected. Labels in the legends refer to direct imaging (blue solid), unpolarized III (red solid), $\hat{\phi}$-polarized III (yellow dashed), $\hat{r}$-polarized III (purple dashed), and the combination of separate $\hat{\phi}$- and $\hat{r}$-polarized III measurements (green dotted)..}
    \label{fig:CRBiso}
\end{figure}

Figure \ref{fig:CRBiso} shows the computed CRBs in the case that both sources are isotropic, as modeled according to the details outlined in Section \ref{section_theory}. The picture is qualitatively similar to those in Fig. \ref{fig:CRBoffangles}. In particular, the $\hat{r}$-polarized III measurement contributes relatively little information across the range of subdiffraction separations. We recently reported an experimental realization of $\hat{\phi}$-polarized III of approximately isotropic emitters \cite{mitchell_quantum-inspired_nodate}. The resolution enhancement achieved there was well worth the price of throwing away the $\hat{r}$-polarized light. 

\section{Conclusion}

We have established the classical and quantum Cram\'{e}r-Rao bounds associated with estimating the separation between a pair of closely spaced, non-interacting, mutually incoherent point sources subject to high-NA collection. Because of the high NA, the scalar approximation is no longer valid and the full dipolar nature of the emission must be considered. We treat two illustrative limiting cases: one in which both emitters have fixed, equal, and known orientations $(\Theta,\Phi)$, and another in which both sources are isotropic. The latter is relevant when single dipole emitters are free to sample all of orientational space over the course of the measurement, or when an ensemble of dipole emitters with uniformly distributed orientations are located at approximately the same position in space such that they can be treated as a single point-like source. In all cases considered we find the QCRB to be much smaller than the CRB of direct imaging at small separations, indicating the existence of a measurement that can obtain far superior source-pair resolution. Parity sorting based on unpolarized III microscopy saturates the QCRB for some special cases of $(\Theta,\Phi)$, as it does within the scalar approximation. However, in the cases of isotropic emitters or general $(\Theta,\Phi)$, the III approach requires polarization filtering in the radial-azimuthal basis to restore super-resolving capabilities. For all orientations except those near $\Theta=0$, the azimuthally polarized component carries sufficient information such that the penalty incurred by throwing away the radially polarized light is not particularly injurious. Unless emitters with $\Theta$ near 0 are of specific interest, throwing away the radially polarized light, or else using it to perform some complementary task like centroid estimation, might provide the most practical solution.

In our endeavor to focus the discussion on the physical phenomenon of interest, we have admittedly considered a simplistic model that implies a certain degree of prior information. It's well-known that uncertainty in  relative brightness \cite{rehacek_multiparameter_2017,rehacek_optimal_2018,prasad_quantum_2020,bonsma-fisher_realistic_2019}, centroid \cite{chrostowski_super-resolution_2017, grace_approaching_2020, parniak_beating_2018}, and number of sources in the scene \cite{bisketzi_quantum_2019,lupo_quantum_2020} will complicate the source-pair resolution problem, and we leave it to future work to fold these ingredients back into the mix. The assumptions implied in the two limiting cases we consider (that either the dipole pair have fixed, known, and equal orientations or else the sources are isotropic) might appear strong at first blush, but they provide useful constraints for the most general case. For one, if the orientations are fixed, known, and \textit{unequal}, then this should only make the resolution problem easier, since unequal orientations imply that additional features of the field (linear polarization, angular distribution) can be leveraged to discriminate the sources. Second, any rotational mobility \cite{lew_rotational_2013} or uncertainty in orientation can modeled by orientational averaging of the appropriate density operator, and the limit of complete rotational mobility or orientational uncertainty coincides with the case of isotropic emitters. Thus the general case should produce trends intermediate to those detailed in our two limiting cases.

For single-parameter estimation problems such as the one considered here, it is known that a measurement saturating the QCRB can be obtained mathematically via orthogonal projection onto the eigenstates of the SLD \cite{braunstein_statistical_1994}. Furthermore, for single-photon states, it is known that any such measurement can be realized algorithmically by a meshwork of Mach-Zehnder interferometers \cite{ReckPRL1994}. Finding a practical measurement scheme that saturates the QCRB, however, often requires some additional imagination and/or empiricism. Since the polarized III comes so close to the QCRB, we aren't particularly motivated to hunt for a more perfect microscope arrangement.

\section{Acknowledgments}
This work was supported by the NSF Science and Technology Center for Quantitative Cell Biology via NSF Grant 2243257. We are grateful to the School of Chemical Sciences' computing department for support and access to the SCS HPC Cluster.

\section{Author Declarations Section}
AID and MPB did the theory. AID and AK performed the computations. AID, AK, and MPB analyzed the data. CSM, DD, and DJD provided supporting data. AID, AK, and MPB wrote the paper.

\section{Data Availability Statement}
Code and data are available upon request.


%
%

%


\bibliography{my_bibliography_aip.bib}

\end{document}